# Free energy barrier for melittin reorientation from a membrane-bound state to a transmembrane state

Sheeba J. Irudayam[a], Tobias Pobandt[b] and Max L. Berkowitz[a*]

[a] *Department of Chemistry, University of North Carolina at Chapel Hill, Chapel Hill, North Carolina, 27599*
[b] *Theory and Bio-Systems, Max Planck Institute of Colloids and Interfaces, Potsdam, Germany, D-14424*

\* Corresponding author. Tel: +1 919 962 1218; Fax: +1 919 962 2388.

E-mail address: isheeba@email.unc.edu (S. J. Irudayam), Tobias.Pobandt@mpikg.mpg.de (T. Pobandt), maxb@unc.edu (M. L. Berkowitz).

**Abstract:**

An important step in a phospholipid membrane pore formation by melittin antimicrobial peptide is a reorientation of the peptide from a surface into a transmembrane conformation. In this work we perform umbrella sampling simulations to calculate the potential of mean force (PMF) for the reorientation of melittin from a surface-bound state to a transmembrane state and provide a molecular level insight into understanding peptide and lipid properties that influence the existence of the free energy barrier. The PMFs were calculated for a peptide to lipid (P/L) ratio of 1/128 and 4/128. We observe that the free energy barrier is reduced when the P/L ratio increased. In addition, we study the cooperative effect; specifically we investigate if the barrier is smaller for a second melittin reorientation, given that another neighboring melittin was already in the transmembrane state. We observe that indeed the barrier of the PMF curve is reduced in this case, thus confirming the presence of a cooperative effect.

**Keywords:**

Melittin reorientation; Umbrella sampling; Bilayer deformations; Antimicrobial peptide action; Cooperative effect.

**Abbreviations:**

S – surface or parallel conformation of the peptide to the bilayer; T – transmembrane or perpendicular conformation of the peptide to the bilayer; PMF – potential of mean force; $N_T$ – number of peptides in the transmembrane conformation.

# 1. Introduction:

Although the mechanism of action of antimicrobial peptides (AMPs) is studied extensively, it is still not well understood and the interpretation of the results obtained, using different experimental techniques is still a subject of many discussions [1-3]. Melittin (MLT) peptide represents an often-studied AMP that serves as an example of a peptide creating toroidal pores in membranes [4-25]. It was observed that at a low peptide to lipid ratio (P/L) melittin adsorbs on a surface of a membrane and assumes a helical conformation with its helical axis parallel to the surface[10]; this state of MLT is called a surface or an S state. As the P/L ratio increases, it was proposed that MLT peptides can reorient and assume a transmembrane conformation, which is called a T state [10]. It was also suggested that peptides in T state may assemble and some of such assemblies may became toroidal pores, but the energetics of how MLT peptides attain their T states and the driving force for the reorientation process is still unknown.

In addition to experimental techniques, computational tools were applied to study MLT action [25-38]. Some of the issues that were investigated through computations were evolution of the structure of pre-formed pores containing MLT peptides[27, 30, 32], the importance of the distribution of charged residues and imperfect amphipathicity in the toroidal pore formation[35, 37]. Atomistic simulations were performed to compare MLT properties in different solvent and membrane environments[28, 38] or to compare MLT to other antimicrobial peptides[25]. Both atomistic and coarse-grained simulations were done to investigate spontaneous pore formation [29, 34, 36]. These studies repeatedly concluded that a perfect pore with all MLT peptides being in a perpendicular orientation does not occur spontaneously on a time scale of simulations (tens of microseconds in coarse-grained simulations). Nevertheless, it was observed that at least one of the peptides adopted a transmembrane or a pseudo-transmembrane orientation in the pore[39]. This suggested that reorientation into a transmembrane state might be an important step in the mechanism of melittin's antimicrobial action.

To understand the details of the structural changes during melittin's reorientation and especially the role of water in this process, we recently performed a computational study using an atomistic description of the system containing MLT peptides embedded into a palmitoyl-oleoyl-phosphatidylcholine (POPC) bilayer [33]. In that work we concentrated on the study of the structural changes in the bilayer, as we forcefully reoriented the N-terminus of MLT from a parallel conformation to a perpendicular conformation. This was done for the P/L ratios of 1/128 and 4/128. Very recently we also performed coarse-grained simulations (using MARTINI force field) on a system containing MLT peptides interacting with phospholipid membrane containing a mixture of zwitterionic and anionic lipids and we saw that at a high P/L ratio (P/L ~ 1/20) MLT can spontaneously create small transient pores in the phospholipid bilayer[39]. Our simulations showed that to create pores some of the MLT peptides had to reorient from a surface state to a transmembrane state.

Although we described previously the structural changes that occur in the lipid membrane during MLT reorientation, the energetic changes were not investigated. Since it is very important to understand thermodynamic and kinetic factors describing MLT reorientation, we present here calculations for the free energy profiles for these processes occurring at different conditions. An important issue in the free energy calculation is the choice of the coordinate(s) for which the free energy curve (surface) is calculated. Most of the recent calculations consider the distance between the z-coordinate (along the bilayer normal) of the center of mass (c.m.) of the peptide and the z-coordinate of the bilayer center to be a proper reaction coordinate[40, 41]. If this coordinate is chosen in calculations, one needs to sample a very large number of possible conformations that may be available to the peptide. To reach some of the conformations may require passage over the barrier and therefore one needs to use methods like replica exchange to explore fully the conformational space. Eventually, one needs to combine replica exchange method with a method such as an umbrella sampling, which requires performance of very large simulations[40]. In our study with melittin we choose a different reaction coordinate: we use as a reaction coordinate the z-coordinate of the distance between the c.m. of three N-terminus residues to the z-coordinate of the bilayer center. This choice of the reaction coordinate was influenced by the knowledge we gained from our coarse-grained simulations: specifically, that melittin peptide reorients in such a way that its N-terminus moves from one leaflet of the bilayer towards the opposing leaflet, while the C-terminus of melittin remains attached to the same leaflet. Since the number of possible conformations the peptide explores in this process is much smaller, we can use just an umbrella sampling method to calculate the free energy change for the reorientation of melittin from a surface state to a transmembrane state.

In the work we present here we calculate the shapes of the free energy curves for two different P/L ratios, a low one when P/L=1/128 and a high one, when P/L=4/128. For the P/L of 4/128 we also perform an additional calculation of the free energy curve when we reorient a MLT peptide in the presence of a second neighboring peptide already reoriented into a T state. As we expected, we find that there is a barrier in the free energy curve for the MLT reorientation, which depends on the P/L ratio and on the states of other peptides. Therefore, the objective of this work is to observe how the barrier heights change with the increase in peptide concentration and to understand if there are any reorientation cooperative effects due to the presence of other peptides in the transmembrane orientation.

## 2. Methodology:

To calculate the free energy of the MLT peptide reorientation from S state to a T state we performed umbrella sampling simulations[42] on three systems: in system A we studied the reorientation of MLT at a P/L ratio 1/128; in system B – reorientation of MLT at a P/L ratio of 4/128 where three peptides remained adsorbed parallel to the bilayer surface (*xy* plane) and the fourth peptide was reoriented; in system C – reorientation of MLT at a P/L ratio of 4/128, but when two peptides were adsorbed parallel to the bilayer surface and a third MLT was placed in a perpendicular orientation spanning the bilayer, while the fourth peptide was reoriented. We considered the reorientation (reaction) coordinate to be the *z*-coordinate of the distance between the center of mass of the first three residues of the N-terminus of the peptide being pulled, and the center of mass of the POPC bilayer. The structural changes in systems A and B were reported in our earlier publication[33]. System C was setup similar to system B, except that two peptides were placed parallel to the bilayer surface, while a third peptide was placed in an orientation perpendicular to the bilayer surface and the N-terminus residues of the fourth peptide were placed at different distances from the center of the bilayer. Windows were equidistantly spaced by 0.2 nm along the reaction coordinate, which extends in an interval from −1.4 nm to 1.4 nm. The bilayer leaflet on which the peptides in the S state were adsorbed was the upper leaflet and in this case the reaction coordinate assumed positive values, from 0 to 1.4 nm. The negative values of distance for the reaction coordinate, from −1.4 to 0 nm, correspond to the peptide N terminus being closer to the lower leaflet of the bilayer, and when the N-terminus reached distances around −1.4 nm we observed that the peptide attained the T state. In Figure 1 we show the MLT peptide with its first three residues next to the N-terminus that represent the pulling group.

An umbrella potential of 200 kJ mol$^{-1}$ nm$^{-2}$ was used for each window, but this force constant was not strong enough to hold the pulling group at certain window positions and additional windows with a higher force constant of 1000 kJ mol$^{-1}$ nm$^{-2}$ were included to sample these regions and enable sufficient overlap between windows. The distances sampled with a higher force constant in the three systems were as follows: for system A (−0.7, −0.6, −0.4, −0.1 and 0.1 nm); system B (−0.6, −0.5, −0.3, 0.0 and 0.3 nm); system C (0.0, 0.2, 0.4 and 0.6 nm). Each window was simulated for 150 ns and the data from the last 100 ns were used for the potential of mean force (PMF) calculation and analysis. Melittin coordinates were taken from the crystal structure (PDB ID: 2MLT). A detailed description of the system setup and the procedure for equilibration are described in reference[33]. In addition to the umbrella sampling simulations of systems with peptides, a pure bilayer system with 128 POPC lipid molecules in a 0.1 M NaCl solution was also simulated for 100 ns as a reference to compare bilayer properties. The simulation parameters and conditions we used here are as listed in reference[33]. The PMFs were evaluated using the g_wham tool in GROMACS[43] and the error analysis was performed using the Bootstrap technique incorporated in the g_wham tool. All the analyses were performed using the GROMACS tools and plotted using Xmgrace and MATLAB®(2010a). Snapshots were visualized using VMD-1.8.6[44].

## 3. Results and Discussion:

*3.1. Insertion Barriers:*

The potentials of mean force (PMF) for the three pulling simulations are given in Figure 2. All three PMFs have a barrier that the system has to cross in order to reorient from a parallel to a perpendicular orientation. The heights of the barriers are as follows: for system A: $13.2 \pm 0.8$ kcal $\text{mol}^{-1}$, for system B: $9.6 \pm 1.9$ kcal $\text{mol}^{-1}$ and for system C: $5.1 \pm 0.8$ kcal $\text{mol}^{-1}$. It can be seen that with an increase in the P/L ratio there is a decrease in the height of the barrier, which is consistent with an idea that the reorientation of the peptide may be slow or even kinetically forbidden at a low P/L, but it becomes possible due to the lowering of the barrier at higher P/L. We also observe that when one MLT is already in a perpendicular orientation, the second MLT has to overcome a reduced barrier of just ~5 kcal $\text{mol}^{-1}$ in order to reorient, indicating a presence of a cooperative effect.

The earlier work, investigating systems A and B, compares the structures of intermediates in the reorientation process, when the N-terminus of the peptide was close to the center of the bilayer[33]. We observed that in system A no continuous water pore was present in the bilayer, while in system B the formation of a water pore stabilized the pseudo-transmembrane state. The distribution of water density from systems A, B and C are given in Figure 3 along with snapshots from each system when the pulling group can be found located near the center of the bilayer. The water density distribution for system C looks very similar to the distribution observed for system B, showing the formation of a water pore near the bilayer center, therefore stabilizing the pseudo-transmembrane state for the second peptide. Thus the presence of water pore could explain the reduction in barrier height for systems B and C compared to system A, but it does not explain why system C has a lower free energy barrier compared to system B.

*3.2. Bilayer Deformations:*

To understand how the material properties of the bilayer influence the barrier height and therefore ability to create pores[13], we studied bilayer deformations. In the case of MLT, it was observed that the thinning of the bilayer upon peptide binding eventually leads to the pore formation [45, 46]. According to Huang et al.[47], peptide binding to the bilayer surface produces stress which favors pore formation; stress causes an increase in the membrane area and the membrane thinning. Here we try to understand the bilayer deformations due to the parallel and perpendicular orientations of the peptide. Figure 4a gives the relative probabilities for finding average positions of MLT atoms in the *xy* plane when the peptide is in a parallel orientation, i.e. when the center of mass (c.m.) of three residues next to the N-terminus of MLT is at distance ~1.4 nm from the bilayer center. Figure 4b shows the deformation of the lipid bilayer in this case. Figures 4c and 4d are similar to Figures 4a and 4b, but they depict the situation when the peptide is in a perpendicular orientation, i.e. when the center of mass of the three residues next to the N-terminus is found at a distance around (−1.4) nm from the bilayer center.

The deformations (Figures 4b and 4d) are defined as the difference in the position of the phosphorous atoms in the upper leaflet of the analyzed window and the average position of the phosphorous atoms in the upper leaflet of a pure bilayer without the peptide. The lesser is the deviation from a pure bilayer, the lighter are the colors on the 2-dimensional surface plot. These distributions were obtained from averages over the last 20 ns of the simulations. When the peptide is in a parallel orientation, its residues next to N-terminus lie between −2 and −1 nm along the $x$-axis and from the Figure 4b we see that this section of the bilayer undergoes more thinning compared to the quadrant of the $xy$-plane where the residues next to the C-terminus are present. This could happen because the C-terminus has four consecutive charged residues, which would prefer to stay at the headgoup-water interface. On the other hand, the N-terminus residues (except the protonated GLY-1) are hydrophobic and would prefer to insert into the headgroup-tail interface, dragging the charged GLY-1 and the headgroup atoms along with them, leading to the thinning of the upper leaflet in this region.

To understand the deformations caused by MLT in transmembrane orientation, we measured the $z$-length of the peptide as the distance between the C-alpha atoms of GLY1 and GLN26 and it is 3.2 nm. The distance between the density peaks of the phosphorous atoms in a pure bilayer is 4 nm; therefore MLT produces a negative mismatch in lengths and induces thinning. Figure 4d shows the resulting thinning of the bilayer around MLT position, while the rest of the leaflet shows minimum perturbation. It should be noted that when MLT is in a perpendicular orientation it also perturbs the headgroup atoms in the lower leaflet. In systems A and B, when all the peptides are in the parallel conformation, the bending of the headgroup atoms does not occur until the N-terminus of the peptide that is pulled reaches the center of the bilayer. At this point the pore is formed facilitated by the creation of water defects that enable the N-terminus to cross the bilayer center and hence reorient. In the case of system C however, since one of the peptides is already in the transmembrane orientation, the headgroups in the lower leaflet are already bent, and therefore it is easier for the snorkeling LYS7 residue to create a water defect in the lower leaflet. This is also reflected by the shift in the pore formation distance for system C (the water pore forms at a distance ~0.4 nm) compared to systems A (~0.1 nm) and B (~0 nm) (see figure 3). Thus, although in both systems B and C a water pore is created which reduces the barrier height compared to system A, the ease with which the water pore is created explains the lower barrier height for system C compared to system B.

Since there is only one MLT peptide in system A, the bilayer deformations could be easily correlated to the parallel and perpendicular orientations of the peptide. The 2-dimensional distributions, like the ones shown in Figure 4, for the deformations in systems B and C are however complicated, because it is difficult to isolate the effect of the S and T orientations; therefore we analyze the average density distributions given in Figure 5. For each system, we compare the densities of the bilayer for the windows at distances of 1.4 nm and −1.4 nm, corresponding, respectively, to the parallel and perpendicular orientations of the peptide being pulled, with the density for a pure bilayer. This comparison shows that for system A the bilayer deformation is minimal. System B experiences the most thinning because at 1.4 nm all four peptides are adsorbed on a membrane surface in an orientation parallel to the surface, inducing a higher stress on the upper leaflet. System C also experiences thinning, but a lesser one than the system B. This could happen because for the window at 1.4 nm only three out of the four peptides are in the parallel orientation and at −1.4 nm only two peptides remain on the surface, which further reduces the local stress on the bilayer.

The area of the bilayer is also a quantity that captures the variation in the bilayer stress caused by the peptide. Table 1 lists the area of the bilayer for the windows at 1.4 nm and at (−1.4) nm for the three systems studied. The areas clearly indicate that the stress on the bilayer is higher when MLT is in a parallel orientation compared to states when MLT is in a perpendicular orientation. For systems B and C, in spite of having the same P/L ratio, the area of the bilayer is smaller when two peptides are in a transmembrane orientation (system C) than when three or all four of them are in a parallel orientation. This release of the stress on the bilayer as peptides adopt a transmembrane orientation could be a contributing factor to the reorientation of the second melittin.

*3.3. Secondary structure:*

It is often assumed that AMPs adopt a helical conformation when they partition into the hydrophobic bilayer core. We monitored the secondary structure of MLT when we pulled it from S to T state. For all three systems we studied, most of the residues remained in an alpha-helical conformation, except the residues near the two termini, which assumed mostly a coil conformation. Recent computational studies[48] on the insertion of hydrophobic peptides have observed that monomeric polyleucine segments of different lengths adopt a helical conformation on the bilayer surface and during the insertion into a bilayer they just reorient their rigid helix, so no peptide conformational entropy change takes place in the process of peptide insertion. However, in the case of MLT, as Figure 6 clearly shows, certain regions of the peptide fold and unfold, as the N-terminus is pulled from the upper to the lower leaflet. A similar observation has been made for other antimicrobial peptides[49]. In simulations the helicity of the peptides very strongly depends on the force fields used in these simulations[50]. Therefore, although our analysis of the helicity does depend on the force field we used, it indicates a possible contribution from the peptide conformational entropy to the free energy change upon reorientation of MLT from S to T state. More detailed quantitative calculations of enthalpic and entropic contributions will be a subject of future work and are not pursued in the work presented here.

## 4. Conclusions:

In this work we calculated the free energy profiles for the reorientation of melittin in a POPC bilayer from a parallel (S state) to a perpendicular (T state) conformation and studied the peptide/lipid interactions that influence this reorientation process. We found that the free energy profiles display barriers and that an increase in the peptide concentration reduces the height of the barrier. We also observed that for a given P/L the presence of other peptides in a T state further reduces the barrier height. The values of the barriers obtained also suggest molecular level insights into understanding the experimental observations and time scales. When the P/L ratio is small, a barrier of ~22 kT is obtained, which can make the reorientation time to be very long. Even if some of the peptides may reorient at a low P/L, the cooperative effect we observed for melittin reorientation cannot play a significant role in the next step of pore construction in this case, since the density of peptides on a surface will be low and the probability of peptides to be located close to each other on a membrane surface will be small. With the increase of the P/L ratio, the barrier for reorientation is lowered and the reorientation time is shortened by at least two orders of magnitude. In addition, at high P/L the probability of finding MLT peptides in close proximity increases, and the cooperative effect will play an important role. The cooperative effect further reduces the value of the barrier and makes it possible for the consequent MLT peptides to reorient in a short time. Our calculation of a free energy for the second peptide is done for a certain distribution of peptides on a surface and therefore can be considered as a conditional free energy. To get an unconditional value for the reorientation free energy profile for the second peptide one needs to perform similar calculations with an ensemble of different initial conditions for peptides, the task requiring vast computational resources. Nevertheless, our computations clearly show the presence of cooperative effect in the reorientation of melittin peptides.

The importance of water defects/pores in stabilizing the reorientation intermediates was discussed in our earlier work where we used a detailed description of atoms in the simulations[33], but we did not study the free energy change and changes in the bilayer. Here, in addition to the calculation of free energy profiles we also studied the correlation between the free energy and the change in the bilayer properties upon the peptide reorientation. We observed that adsorbed to the surface MLT peptides increase the bilayer surface area and make the bilayer thinner. As the P/L ratio increases, the bilayer area increases, but the reorientation of the peptide reduces the area. If the surface area is a measure of the stress experienced by the bilayer, its reduction upon the peptide reorientation indicates the release of the stress; as previously proposed this could be a driving force for the reorientation of MLT[46].

Although the results of our free energy calculations depend strongly on the force fields (as in many other similar by spirit calculations), we believe that our results clearly show that MLT undergoes reorientation when the P/L ratio is large and that a cooperative effect plays a significant role in the pore creation by MLT peptides in lipid membranes.

**Acknowledgements**

This work was supported by the National Science Foundation under grant MCB-0950280. Tobias Pobandt was funded by the Deutsche Forschungsgemeinschaft (DFG) via the IRTG 1524. We also thank Dr. Volker Knecht for helpful discussions.

**Tables:**

Table 1. Areas of the bilayer for a system without peptides and for the bilayers in the windows at 1.4 nm and −1.4 nm of systems A, B and C corresponding to the parallel and perpendicular orientations respectively of the peptide being pulled. $N_S$ and $N_T$ are the number of peptides in the parallel and perpendicular orientations in each window.

| System | $N_S$ | $N_T$ | Area / nm$^2$ |
|---|---|---|---|
| Pure bilayer | - | - | 40.52 ± 0.23 |
| A | 1 | 0 | 41.47 ± 0.28 |
| A | 0 | 1 | 41.36 ± 0.11 |
| B | 4 | 0 | 44.30 ± 0.20 |
| B | 3 | 1 | 43.78 ± 0.13 |
| C | 3 | 1 | 43.49 ± 0.08 |
| C | 2 | 2 | 42.29 ± 0.04 |

**Figure Captions:**

Figure 1. A snapshot of the system with the P/L = 1/128 when the N-terminus residues are around 1 nm from the bilayer center (dashed line). The reaction coordinate scale in nm is shown on the left, while and residues of the peptide pulling group are inside the grey circle. Phosphorous atoms of the bilayer are shown as orange spheres (the rest of the bilayer atoms are not shown for clarity) and water as blue dots. The peptide is colored by its residue type (hydrophobic: yellow, hydrophilic: green, charged: blue).

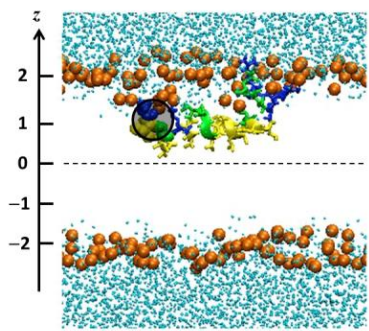

Figure 2. PMFs for the reorientation of melittin in system A (a), system B (b) and system C (c).

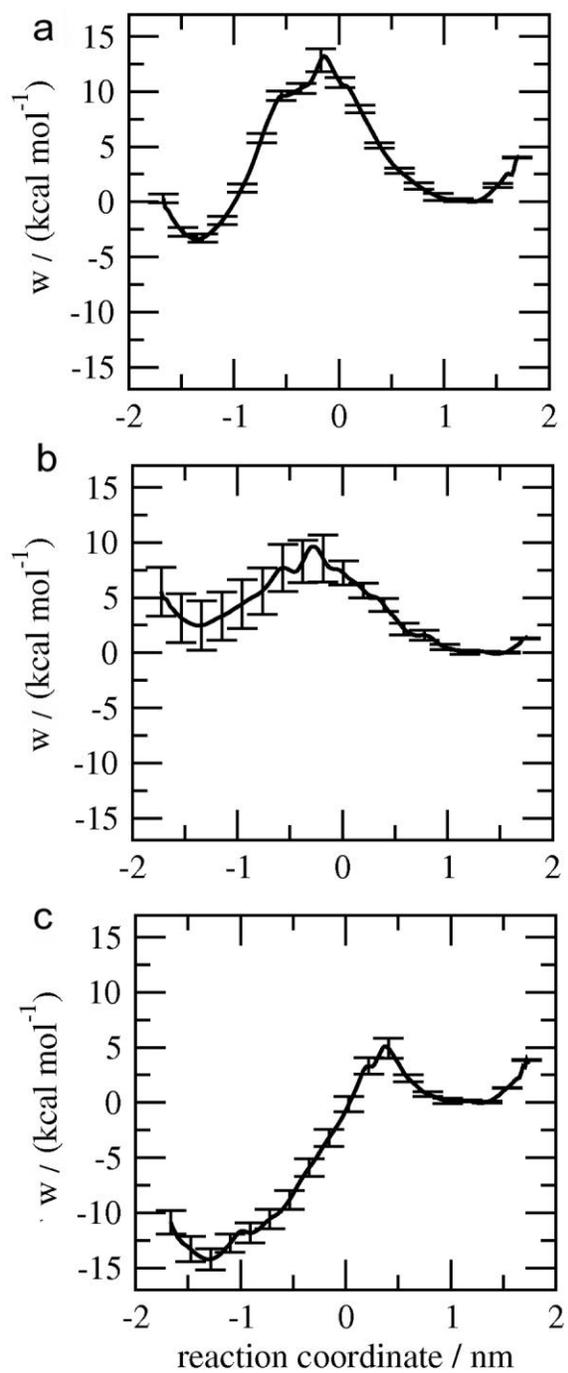

Figure 3. The distribution of number density of water (nm$^{-3}$) as a function of the reaction coordinate and the distance from the bilayer center in system A (a), system B (b) and system C (c) along with a snapshot for the corresponding system shown below each distribution. The snapshots represent the systems when the pulling group (shown as VDW spheres in magenta) is at regions near the center of mass of the bilayer. The description of the colors for the snapshots is the same as in Figure 1.

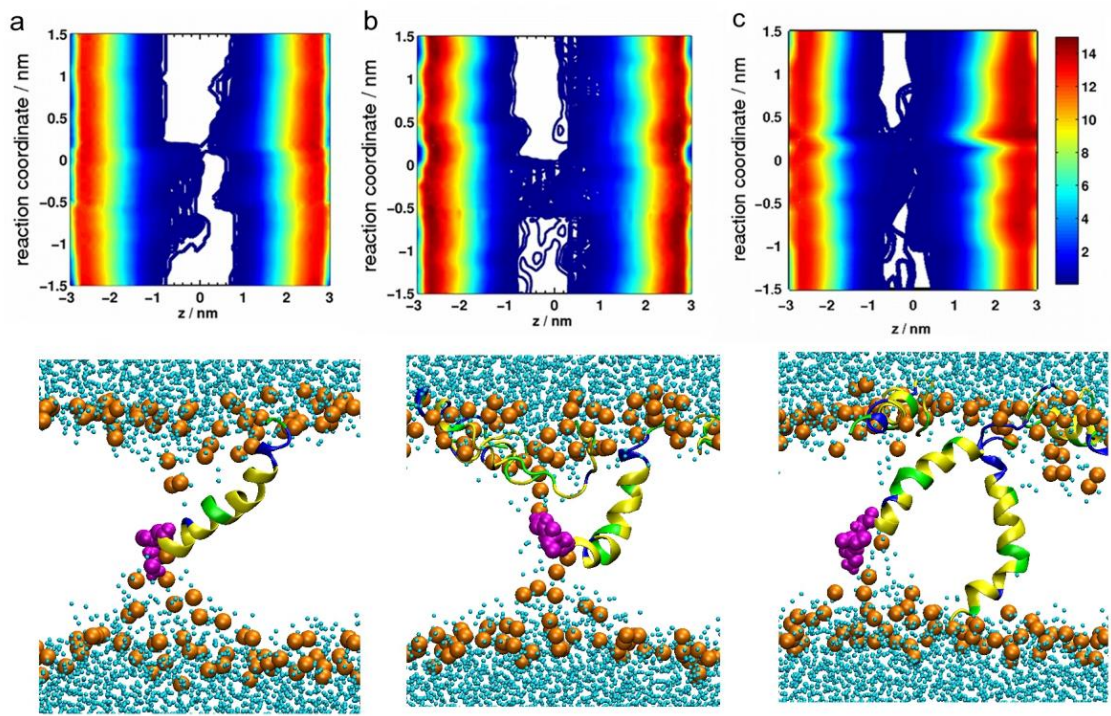

Figure 4. a) The probability of finding a peptide atom in the *xy* plane of system A averaged over the last 20 ns of the window at 1.4 nm. b) Bilayer deformation (nm) in system A for the window at 1.4 nm. c) The probability of finding a peptide atom in the *xy* plane of system A for the window at −1.4 nm. d) Bilayer deformation in system A for the window at −1.4 nm.

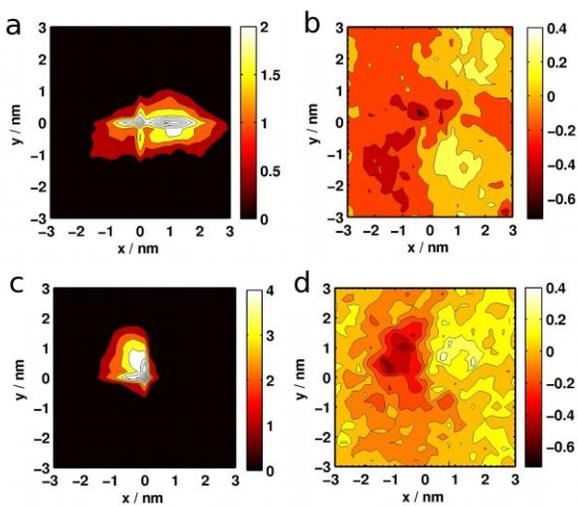

Figure 5. Mass density distributions of a bilayer without peptides (grey) and the bilayer in windows at 1.4 nm (solid black) and (−1.4) nm (dashed black) for system A (a), system B (b) and system C (c).

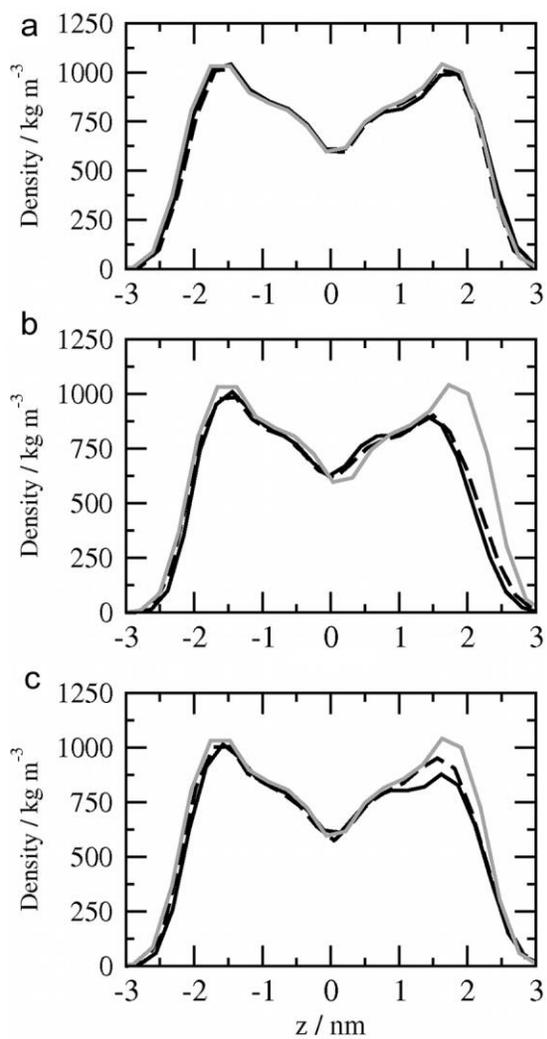

Figure 6. The secondary structure of each residue of the peptide being pulled as a function of the reaction coordinate for system A (a), system B (b) and system C (c). Colors for the secondary structures are as follows: coil – red; α-helix – blue; bend/turn – yellow; 5-helix/3-helix/β-sheet – green.

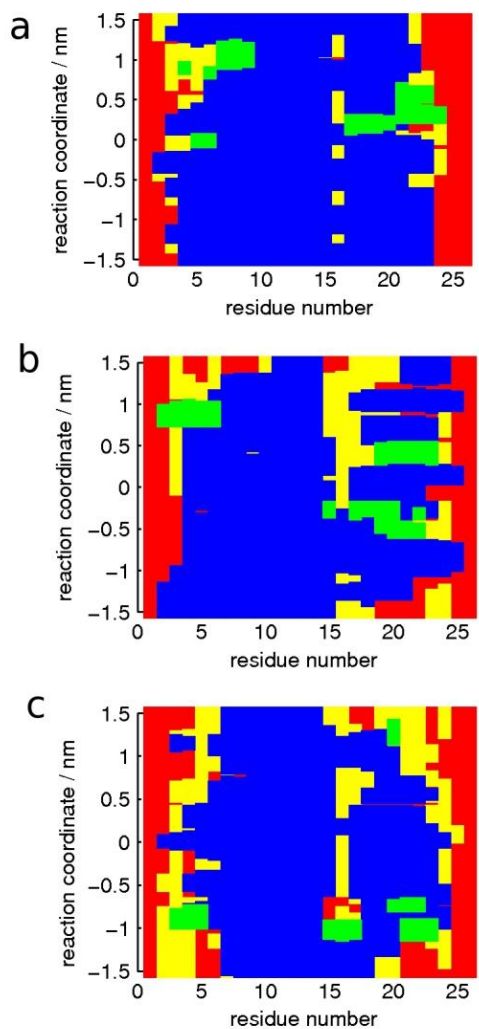